\documentclass[9pt,conference]{IEEEtran}

\usepackage[preprint]{waspaa25}

\usepackage{bm} 

\title{Post-Training Quantization for Audio Diffusion Transformers}

\name{Tanmay Khandelwal$^{1,2}$,
      Magdalena Fuentes$^{2,3}$}
\address{$^{1}$Courant Institute of Mathematical Sciences, New York University, NY, USA \\
$^{2}$MARL, New York University, NY, USA\\
$^{3}$IDM, New York University, NY, USA
}

\begin{document}

\maketitle

\begin{abstract}
Diffusion Transformers (DiTs) enable high-quality audio synthesis but are often computationally intensive and require substantial storage, which limits their practical deployment. In this paper, we present a comprehensive evaluation of post-training quantization (PTQ) techniques for audio DiTs, analyzing the trade-offs between static and dynamic quantization schemes. We explore two practical extensions (1) a denoising-timestep-aware smoothing method that adapts quantization scales per-input-channel and timestep to mitigate activation outliers, and (2) a lightweight low-rank adapter (LoRA)-based branch derived from singular value decomposition (SVD) to compensate for residual weight errors. Using Stable Audio Open we benchmark W8A8 and W4A8 configurations across objective metrics and human perceptual ratings. Our results show that dynamic quantization preserves fidelity even at lower precision, while static methods remain competitive with lower latency. Overall, our findings show that low-precision DiTs can retain high-fidelity generation while reducing memory usage by up to 79\%.
\end{abstract}

\section{Introduction}
\label{sec:intro}

Diffusion models are a powerful type of generative model that excel at creating high-quality outputs in areas like audio generation \cite{floresgarcia2024sketch2sound,pmlr-v202-liu23f}. They are increasingly being adopted in music production \cite{controllable-music, huang2023noise2music, 10.1145/3677996.3678289} and sound design \cite{zhu2023edmsoundspectrogrambaseddiffusion}. Compared to generative adversarial networks (GANs) and variational autoencoders (VAEs), diffusion models have more stable training and avoid issues like model collapse, making them a great choice for audio generation tasks. Diffusion transformers (DiTs) outperform traditional diffusion models with UNet backbones in both performance and flexibility \cite{peebles2023scalablediffusionmodelstransformers, bao2023worthwordsvitbackbone}. The hierarchical convolutional structure of UNet models presents scalability challenges, limiting their effectiveness in handling complex tasks like audio generation \cite{ding2025posttrainingquantizationdiffusiontransformer}. In contrast, DiTs \cite{10.5555/3692070.3692575, tian2025audioxdiffusiontransformeranythingtoaudio} leverage transformer architectures to better capture long-range temporal dependencies and intricate spectral patterns that are critical in audio generation. This makes them great for tasks like generating realistic instrument sounds \cite{article}, smooth soundscapes\cite{immerse-diffusion-generative}, and natural-sounding speech \cite{park2024dex} with expressive tones. Models like Stable Audio \cite{10.5555/3692070.3692575} show how these systems can create high-quality audio clips with consistent timing and sound detail.

Despite their success across various generative tasks, DiTs face significant challenges due to their high computational requirements and increased storage demands \cite{lou2024tokencachingdiffusiontransformer,ma2023deepcacheacceleratingdiffusionmodels}. To address this, researchers have turned to model quantization, which reduces computation and memory demands by using lower bitwidths for weights and activations. Among these techniques, post-training quantization (PTQ) stands out as a practical and straightforward approach \cite{zhang2023posttrainingquantizationneuralnetworks}. Unlike quantization-aware training (QAT), which requires retraining the entire model, PTQ uses a small dataset for quick calibration to adjust scale factors and minimize quantization errors. This makes PTQ particularly suitable for quantizing DiTs from 32-bit floating-point weights into 8-bit or 4-bit integers without the need for extensive computational resources. PTQ can also convert the
activations (e.g. the input of a linear layer) from 32-bit float
numbers to 8-bit integers. As a result, the matrix multiplications of both attention modules and linear layers could take place in
the low-precision integer field, thus accelerating the inference process and reducing the memory footprint.

While most quantization research to date has focused on UNet-based diffusion models \cite{Li_2023_ICCV, Zhan_2025_ICLR}—particularly in text-to-audio generation tasks \cite{Vora2024PTQ4ADMPQ}—transformer-based diffusion models such as DiTs remain largely underexplored in the audio domain. This gap is notable given DiT's superior performance in audio generation \cite{10.5555/3692070.3692575}. Most PTQ methods for diffusion models rely on fixed-point quantization, which can introduce significant errors at lower precision, resulting in performance degradation. When these methods are applied to DiTs, two major challenges arise. First, certain channels within the model—often referred to as salient channels—can exhibit extremely large or small values compared to others \cite{Wu2024PTQ4DiT}. This imbalance disrupts uniform scaling, causing substantial quantization errors. Second, the distribution of activations in DiTs changes significantly across different timesteps of the diffusion process. Early timesteps are dominated by noise \cite{Wu2024PTQ4DiT}, while later timesteps focus on refining fine-grained audio details, resulting in highly variable activation ranges throughout inference. As a result, a single, static quantization range is often insufficient to accommodate these variations, leading to cumulative errors and degraded generation quality. Recent work has begun to address quantization challenges, particularly for image generation. These methods target high-activation layers \cite{Vora2024PTQ4ADMPQ} and address activation variability via techniques like channel-wise salience balancing \cite{Wu2024PTQ4DiT}. Building on this, PTQ4DM \cite{shang2023ptqdm} uses timestep-aware calibration, Q-Diffusion \cite{Li_2023_ICCV} introduces split shortcuts for 4-bit quantization, APQ-DM \cite{wang2024accurateposttrainingquantizationdiffusion} applies group-wise rounding, and PTQD \cite{he2023ptqdaccurateposttrainingquantization} adds variance correction for mixed precision. Recent advances include SVDQuant’s \cite{Li2024SVDQuant} low-rank outlier suppression and DiTAS’s \cite{Dong2024DiTAS} layer-wise grid search strategy with temporal smoothing. Despite these advances, there remains limited insight into how DiTs behave specifically in audio generation tasks, whether the same issues arise, and how effectively these models can be quantized.


In this paper, we conduct a comprehensive study of PTQ strategies for audio DiTs. We analyze the behavior of a widely used audio generation DiT model (i.e, we look at activation and weights ranges and outliers), and we introduce two practical extensions tailored to audio DiTs. First, denoising-timestep-aware smoothing strategy based on SmoothQuant \cite{yao2022smoothquant}, which scales activations and weights individually for each timestep and channel, addressing the dynamic activation distributions inherent in diffusion models. Secondly, to mitigate degradation in generation performance, we assess integrating low-rank adaptation (LoRA) \cite{hu2021loralowrankadaptationlarge} modules into the quantized weights of the DiT model. Specifically, we apply singular value decomposition (SVD) to the smoothed and quantized weight matrices, decomposing them into a low-rank component and a residual. This decomposition allows us to compensate for quantization errors by isolating the residuals into trainable low-rank approximations. We investigate the effects of each technique, individually and in combination, across static and dynamic quantization regimes. Our results provide insights into which configurations best preserve the generation quality of audio DiTs, as measured by both objective metrics and human evaluations.



\section{Methodology}
\label{sec:methodology}

For our analysis, we chose Stable Audio Open \cite{evans2024stableaudioopen} because it is fully open-source and provides open access to the model’s weights. Moreover, its architecture combines an autoencoder, T5-based text conditioning, and transformer-based diffusion, which is representative of modern DiTs for audio generation. Finally, the model is optimized for consumer GPUs, has strong community support, and has reproducible computational benchmarks, thus ideal for our study. Same as Stable Audio Open, we use AudioCaps \cite{kim-etal-2019-audiocaps} as benchmark.

\begin{figure}[bht]
    \centering
    \includegraphics[width=\linewidth]{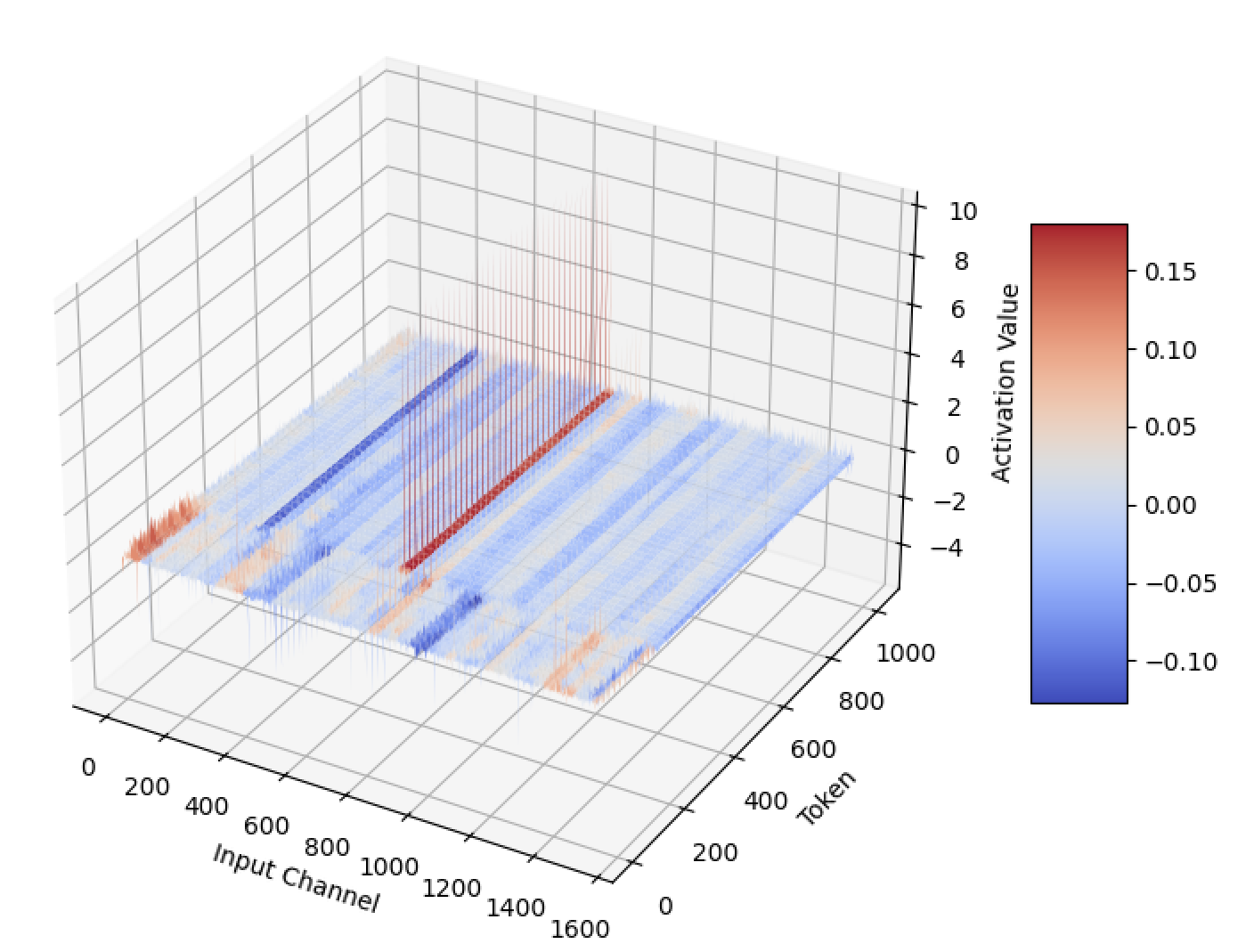}
\caption{Activation map at denoising timestep 50 for DiT Block 24, showing activation values across tokens and input channels.}
\label{fig:activation}\end{figure}

\begin{figure}[bht]
    \centering
    \includegraphics[width=\linewidth]{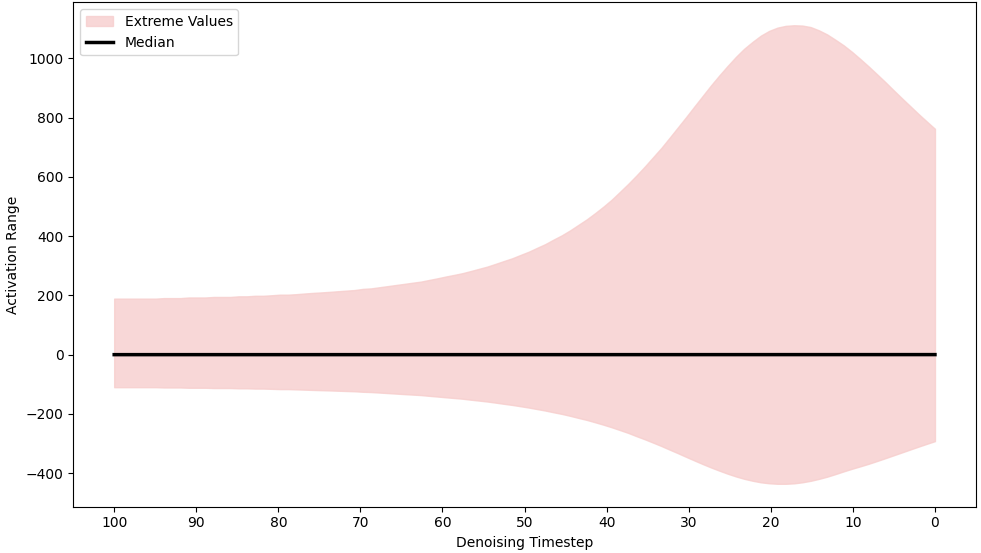}
    \caption{Visualization of input activation range across denoising timesteps (100~$\rightarrow$~0) for Block 1. The shaded region represents the full activation span (min to max), while the solid line denotes the median activation. As denoising progresses, the range of activations increases significantly, highlighting the emergence of outliers in later steps.}
    \label{fig:model}
\end{figure}

Most DiTs, including Stable Audio's, are constructed from stacks of transformer blocks, each comprising self-attention layers and multilayer perceptron (MLP) modules \cite{peebles2023scalablediffusionmodelstransformers, vaswani2023attentionneed}. Within these blocks, both the feed-forward networks (FFNs) and the query-key-value (QKV) projection layers of self-attention are major contributors to computational cost. FFNs alone account for over 60\% of model parameters and up to 70\% of total FLOPs. Similarly, the QKV projections in self-attention require large linear transformations to compute the query, key, and value representations for each token, further increasing the computational and memory demands. As a result, both FFNs and QKV layers are critical targets for PTQ. We start by analyzing the input activations of both FFNs and QKV projections using forward passes on randomly selected prompts from the validation set, recording per-channel activation ranges to understand how activation values behave in audio DiTs. Figures \ref{fig:activation} and \ref{fig:model} provide visual intuition for the design choices in our study.


 First, Figure \ref{fig:activation} shows a 3D visualization of the activation distributions at time step 50 for DiT Block 24, plotting activation values across both tokens and input channels. During our analysis, one major issue we observed was the large variation in activation values across input channels, particularly in the QKV projections of the self-attention layers and the FFN layers. The vertical spikes (both in red and blue, indicating positive and negative values) make clear that certain input channels yield significantly larger magnitudes—sometimes extreme outliers, while others hover near zero. As a result, channels with more moderate activations would suffer from elevated quantization error. Although quantization is typically performed on output channels as it is hardware-efficient, we observed that the uneven activation patterns across these input channels, together with the presence of large outliers, would severely skew the quantization parameters. Motivated by this and following insights from prior work \cite{Dong2024DiTAS}, we instead quantize activations taking into account input channels.


Second, we observed from Figure \ref{fig:model} how the activation values expand during the denoising process, with the horizontal axis representing the denoising timesteps and the vertical axis capturing the range of activation values. As the timesteps advance, the distribution widens and occasionally spans extreme magnitudes. Static quantization methods, which are designed around fixed activation ranges, would result in amplified errors at later timesteps.

Our objective is to tackle outliers in the activations across input channels, and the varying ranges of those activations over timesteps. For that we adapt SmoothQuant \cite{yao2022smoothquant}, usually applied in the context of large language models (LLMs), and introduce a per-input-channel, time-aware smoothing factor to reduce the impact of activation outliers. By storing the maximum activation values for each channel and timestep during the denoising process, we dynamically adjust quantization parameters to account for temporal and channel-wise variation.

Our extension is as follows: Consider a linear layer where \( \mathbf{X}^{(t)} \in \mathbb{R}^{k \times n} \) denotes the activation matrix at timestep \( t \), with \( k \) channels and \( n \) elements per channel. Let \( \mathbf{W} \) be the corresponding weight matrix.

For each channel \( j \in \{1, \ldots, k\} \), we record the maximum absolute activation:
\begin{equation}
\mathbf{X}_{\text{absmax},j}^{(t)} = \max\left(|\mathbf{X}_j^{(t)}|\right),
\label{eq:x_absmax}
\end{equation}
and the corresponding maximum absolute weight:
\begin{equation}
\mathbf{W}_{\text{absmax},j} = \max\left(|\mathbf{W}_j|\right).
\label{eq:w_absmax}
\end{equation}

Using these values, we define a per-channel smoothing factor:
\begin{equation}
\mathbf{s}_j^{(t)} = \frac{\left(\mathbf{X}_{\text{absmax},j}^{(t)}\right)^{\alpha}}{\left(\mathbf{W}_{\text{absmax},j}\right)^{1-\alpha}}, \quad \alpha \in [0, 1],
\label{eq:s_factor} 
\end{equation}
which balances the influence of activations and weights. A larger \( \alpha \) results in stronger attenuation of large activation values, while a smaller \( \alpha \) emphasizes weight scaling.

This smoothing is implemented by rescaling both activations and weights:
\begin{equation}
\hat{\mathbf{X}}_j^{(t)} = \frac{\mathbf{X}_j^{(t)}}{\mathbf{s}_j^{(t)}}, \qquad
\hat{\mathbf{W}}_j = \mathbf{W}_j \cdot \mathbf{s}_j^{(t)},
\label{eq:smoothing_scale}
\end{equation}
such that the resulting linear transformation remains algebraically identical:
\begin{equation}
\mathbf{Y} = \hat{\mathbf{X}}^{(t)} \hat{\mathbf{W}} = \left(\frac{\mathbf{X}^{(t)}}{\mathbf{s}^{(t)}}\right) \left(\mathbf{W} \cdot \mathbf{s}^{(t)}\right) = \mathbf{X}^{(t)} \mathbf{W}.
\label{eq:invariance}
\end{equation}

\begin{figure}[bht]
\centering
\includegraphics[width=\linewidth]{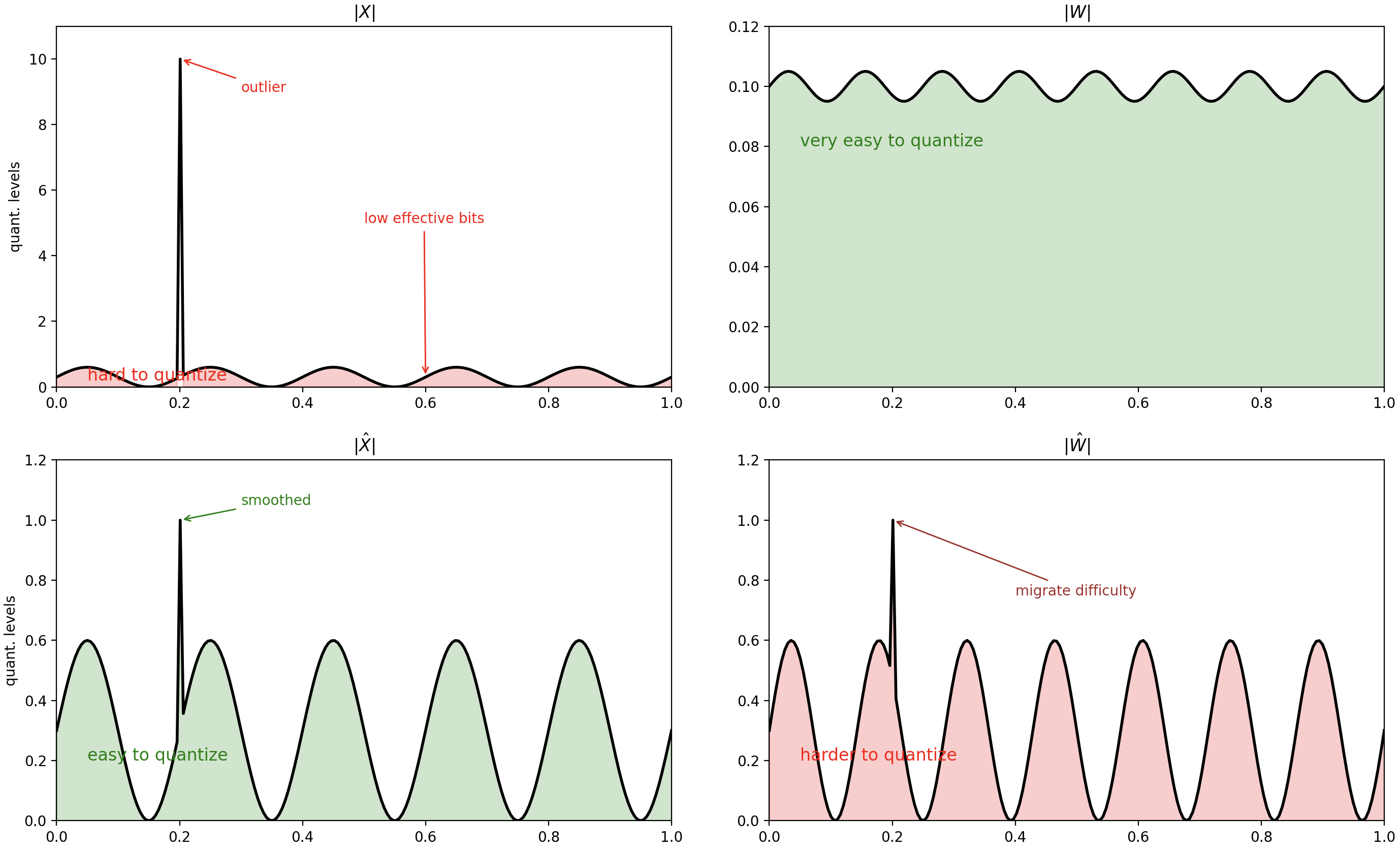}
\caption{Visualizing “easy vs. hard to quantize” regions and outliers. The spikes in activations (“outliers”) lead to low effective bits for other channels, whereas flatter distributions (“smoothed”) are more amenable to quantization.}
\label{fig:uploaded}
\end{figure}

Figure~\ref{fig:uploaded} shows the intuition behind this. The top-left panel shows the absolute activation values \( |\mathbf{X}| \) before smoothing. Here, a single large outlier dominates the range, forcing the quantizer to reserve most of its dynamic range for rare, extreme values. This leads to low effective bits for the remaining, more common activation values, making them difficult to quantize precisely. Meanwhile, the top-right panel shows the corresponding weight distribution \( |\mathbf{W}| \), which is smoother with fewer outliers and thus easier to quantize.

To address this imbalance, we compute a smoothing factor that rebalances the dynamic ranges between activations and weights, effectively reducing the impact of outliers during quantization. We call this method SmoothQuant Dynamic (SQD).


Figure~\ref{fig:uploaded} (bottom) further illustrates how Eq.~\eqref{eq:s_factor} flattens sharp activation peaks, yielding distributions that are easier to quantize. While smoothing makes \( \hat{\mathbf{W}} \) more quantization-friendly, the finite-precision representation still introduces residual error. To mitigate this, we introduce a 16-bit low-rank branch. Intuitively, this low-rank branch is aimed at capturing the most important components of the quantization error, and correct them. The transformed weight matrix is decomposed into \( \mathbf{A} \mathbf{B}^{\top} \). Compared to direct 4/8-bit quantization, we first compute the low-rank branch in 16-bit precision and then approximate the residual in 4-bit or 8-bit quantization. As a result, the additional parameters and computational overhead for the low-rank branch are negligible. To find the low-rank branch, we first compute the residual:
\begin{equation}
\mathbf{E} = \mathbf{W} - \hat{\mathbf{W}}, \qquad \mathbf{E} \in \mathbb{R}^{k \times m},
\label{eq:residual}
\end{equation}

then model the structured component of \( \mathbf{E} \) using truncated SVD:
\begin{equation}
\mathbf{E} \approx \mathbf{U}_r \bm{\Sigma}_r \mathbf{V}_r^\top, \qquad r \ll \min(k, m),
\label{eq:svd}
\end{equation}
where \( \mathbf{U}_r \in \mathbb{R}^{k \times r} \), \( \bm{\Sigma}_r = \mathrm{diag}(\sigma_1, \dots, \sigma_r) \), and \( \mathbf{V}_r \in \mathbb{R}^{m \times r} \).

We then define:
\begin{equation}
\mathbf{A} = \mathbf{U}_r \bm{\Sigma}_r^{1/2}, \qquad \mathbf{B} = \mathbf{V}_r \bm{\Sigma}_r^{1/2},
\label{eq:lora_ab}
\end{equation}
so that \( \mathbf{A} \mathbf{B}^\top = \mathbf{U}_r \bm{\Sigma}_r \mathbf{V}_r^\top \). Both \( \mathbf{A} \) and \( \mathbf{B} \) are kept in FP16.

The final weights used at inference time are:
\begin{equation}
\tilde{\mathbf{W}} = \underbrace{\hat{\mathbf{W}}}_{\text{INT8/INT4 core}} + \underbrace{\mathbf{A} \mathbf{B}^\top}_{\text{FP16 adaptor}}.
\label{eq:compensated_w}
\end{equation}

Given an INT8/INT4-quantized input \( \left\lfloor \mathbf{X}^{(t)} / \mathbf{x}_{\text{scale}} \right\rceil \), matrix multiplication with \( \tilde{\mathbf{W}} \) is accumulated in FP16 or FP32 to preserve numerical precision.\footnote{Note that this accumulation in FP16/FP32 does not increase the model’s memory footprint, which aligns with the primary goal of our work. While executing all operations in lower precision could further accelerate computation, this is beyond the scope of this paper and left for future work.}

We also implement a lightweight variant called SmoothQuant Static (SQS). Unlike SQD, which adapts scales dynamically per timestep, this variant applies static quantization to both weights and activations based on precomputed statistics. During calibration, each layer tracks the per-input-channel running minimum and maximum of activations across denoising steps. After collecting these statistics, we compute a single global maximum per channel to derive the SmoothQuant scale and fold it into the weights. Activations are then quantized using the global min/max range, producing fixed scaling parameters. This requires no runtime adaptation or fine-tuning, making SQS efficient and low-latency.

\section{Experimental design}
This section details the experimental process for our quantized Stable Audio model on the audio generation task. Our methodology closely follows the evaluation protocol described in the original Stable Audio Open paper~\cite{evans2024stableaudioopen}. We use the pre-trained Stable Audio Open model as our full-precision baseline, operating at a 44.1 kHz sampling rate and generating 10-second audio clips.

For audio generation, we employ the DPM-Solver++ sampler with 100 steps, using classifier-free guidance (CFG) set to 7.0 to enhance output quality. Noise levels are managed with $\sigma_{\text{min}} = 0.3$ and $\sigma_{\text{max}} = 500$. The model sourced from Hugging Face, serves as the foundation for our experiments, upon which we apply various quantization techniques.

The evaluation is conducted using the AudioCaps evaluation dataset~\cite{kim2019audiocaps}, which originally contains 979 YouTube audio segments, each paired with multiple captions. After filtering out inaccessible files, we retain 881 audio segments and 4,875 corresponding captions. These captions are used to generate 4,875 audio clips, mirroring the procedure in the Stable Audio Open paper. All experiments are performed on a single NVIDIA A100 GPU. To ensure format compatibility, the audio is peak-normalized, clipped, and converted to 16-bit PCM.

To maintain comparability with Stable Audio Open, we use three established evaluation metrics to thoroughly assess the quality and relevance of audio generated by our quantized Stable Audio model. The first metric, FD\textsubscript{openl3}, compares the feature distributions of generated and reference audio. Lower FD\textsubscript{openl3} scores indicate that the generated audio closely resembles real audio, reflecting high fidelity. The second metric, KL\textsubscript{passt}, measures semantic similarity by comparing distributions of audio tags predicted by a pre-trained tagger. A lower KL\textsubscript{passt} score means the generated audio captures the same semantic content as the reference, indicating strong alignment in meaning and content. The third metric, CLAP\textsubscript{score}, evaluates how well the generated audio matches the provided text prompt by comparing embeddings of the audio and its caption. A higher CLAP\textsubscript{score} shows that the generated audio accurately reflects the intent and details of the input text.

To assess model efficiency, we compare the size of the model before and after applying quantization methods such as SmoothQuant and LoRA. Model size is measured by saving the state dictionary and recording the file size, with the original full-precision model occupying approximately 4,854 MB. We focus on two quantization configurations: W8A8 (8-bit weights and activations) and W4A8 (4-bit weights, 8-bit activations).


For PTQ, we use a calibration set of 50 randomly selected prompts. This set is used both for SmoothQuant calibration with the hyperparameter $\alpha$ set to 0.5 and for computing the SVD of the LoRA components. Our implementation applies per-output-channel symmetric quantization for weights and per-input-channel symmetric quantization for activations. Experiments are conducted using both W8A8 and W4A8 configurations to systematically evaluate the trade-offs between compression and generation quality.

\section{Results and discussion}

We establish the baseline using the original, full-precision model. 
During preliminary experiments, we observed that evaluation metrics varied 
substantially based on the random seed, often diverging from the originally reported 
values in earlier studies. To ensure consistency and fairness, we systematically 
tested multiple seeds and ultimately selected {seed = 1000}, which yielded results close to those reported and high-quality generations. Our full-precision results 
achieve a CLAP Score of 0.3009, KL\textsubscript{passt} of 2.17, and 
FDopenl3 of 87.02,  and a best‑case generation
latency of $\sim11.3$\,s. We use this as the baseline, but still include results reported in the original paper in our table.

\begin{table}[hb]
\centering
\caption{Performance comparison of full-precision and quantized Stable Audio models using SmoothQuant and LoRA for both dynamic (i.e. channel- and step-dependent) and static cases (i.e. single value for all channels and steps). SQD = SmoothQuant Dynamic; SQS = SmoothQuant Static. LoRA denotes low-rank adaptation. $\uparrow$ indicates higher is better; $\downarrow$ indicates lower is better. Best results in \textbf{bold}, second best \underline{underlined}.}
\label{tab:quant_results}
\resizebox{\columnwidth}{!}{%
\begin{tabular}{llcccc}
\toprule
\textbf{Precision} & \textbf{Variant} & \textbf{CLAP} $\uparrow$ & \textbf{KL\textsubscript{passt}} $\downarrow$ & \textbf{FD\textsubscript{openl3}} $\downarrow$ & \textbf{Size (GB)} $\downarrow$\\
\midrule
FP32   & Reported     & 0.2900 & 2.14  & \textbf{78.24}  & --   \\
FP32   & Baseline     & 0.3009 & 2.17  & 87.02  & 4.85 \\
\midrule
W8A8   & SQD          & \underline{0.3021} & 2.158 & 86.35  & \underline{1.65} \\
W8A8   & SQS          & 0.2934 & 2.144 & \underline{80.57}  & \underline{1.65} \\
W8A8   & SQD+LoRA     & \textbf{0.3033} & 2.153 & 85.70  & 1.71 \\
\midrule
W4A4   & SQD          & 0.2901 & \textbf{2.039} & 82.57  & \textbf{1.03} \\
W4A4   & SQS          & 0.2014 & 2.780 & 224.7  & \textbf{1.03} \\
W4A4   & SQD+LoRA     & 0.2829 & \underline{2.096} & 85.85  & 1.17 \\
\bottomrule
\end{tabular}%
}
\end{table}
    \vspace{-3mm}
\medskip
We evaluate three quantization strategies under two precision settings: W8A8 and W4A8. The quantization strategies are SmoothQuant Dynamic (SQD) with and without low-rank adaptation (+LoRA), and  SmoothQuant Static (SQS). Please see Section \ref{sec:methodology} for details about these methods. Results are shown in Table \ref{tab:quant_results}. 

We found that SQD models closely match or even surpass our full-precision baseline across objective metrics (CLAP, KL\textsubscript{passt}, and 
FD\textsubscript{openl3}) for the two precision configurations. This shows that dynamic calibration effectively handles activation outliers at each timestep, with minimal performance loss after quantization.

\medskip

For W8A8, LoRA consistently boosted 
metrics, narrowing any remaining gap to the FP32 baseline. At 
W4A8, however, LoRA did not yield consistent improvements. Given that the quantization error is often too severe in this setting, especially in the presence of activation outliers, a low-rank additive correction (like LoRA) falls short.

We noted that the static approach performs competitively at W8A8, offering a strong trade-off between simplicity and quality. However, in the more aggressive W4A8 setting, static quantization results in significant degradation. This suggests that in more aggressive quantization settings, there are bigger advantages to adjusting dynamically to activation outliers.

\medskip

A key limitation of our SQD approach compared to SQS is its slower inference speed ($\sim35.6$\,s vs. $\sim11.6$\,s), which is primarily due to the overhead of maintaining scaling factors that are specific to each timestep and input channel, as well as the need for dynamic computation during inference. This can be mitigated through caching, pruning unused quantization paths, or integrating fast integer-aware operators—remains, and remains an avenue for future work.

We conducted a subjective evaluation to complement our objective metrics. Specifically, we compared our best-performing model (SQD + LoRA) with the fastest configuration (SQS), both in W8A8 precision, alongside the original full-precision baseline. To ensure diversity in auditory content, we selected five prompts spanning various sound classes and constructed a 15-question survey by randomly sampling these prompts across the three model variants. The survey was distributed to 20 participants. The participants were asked to rate each audio sample on a 1–5 scale based on perceived alignment with the input prompt. These qualitative judgments were then aggregated into a single composite score per model to quantify perceptual performance.

\begin{figure}[bt]
    \centering
    \includegraphics[width=0.8\linewidth]{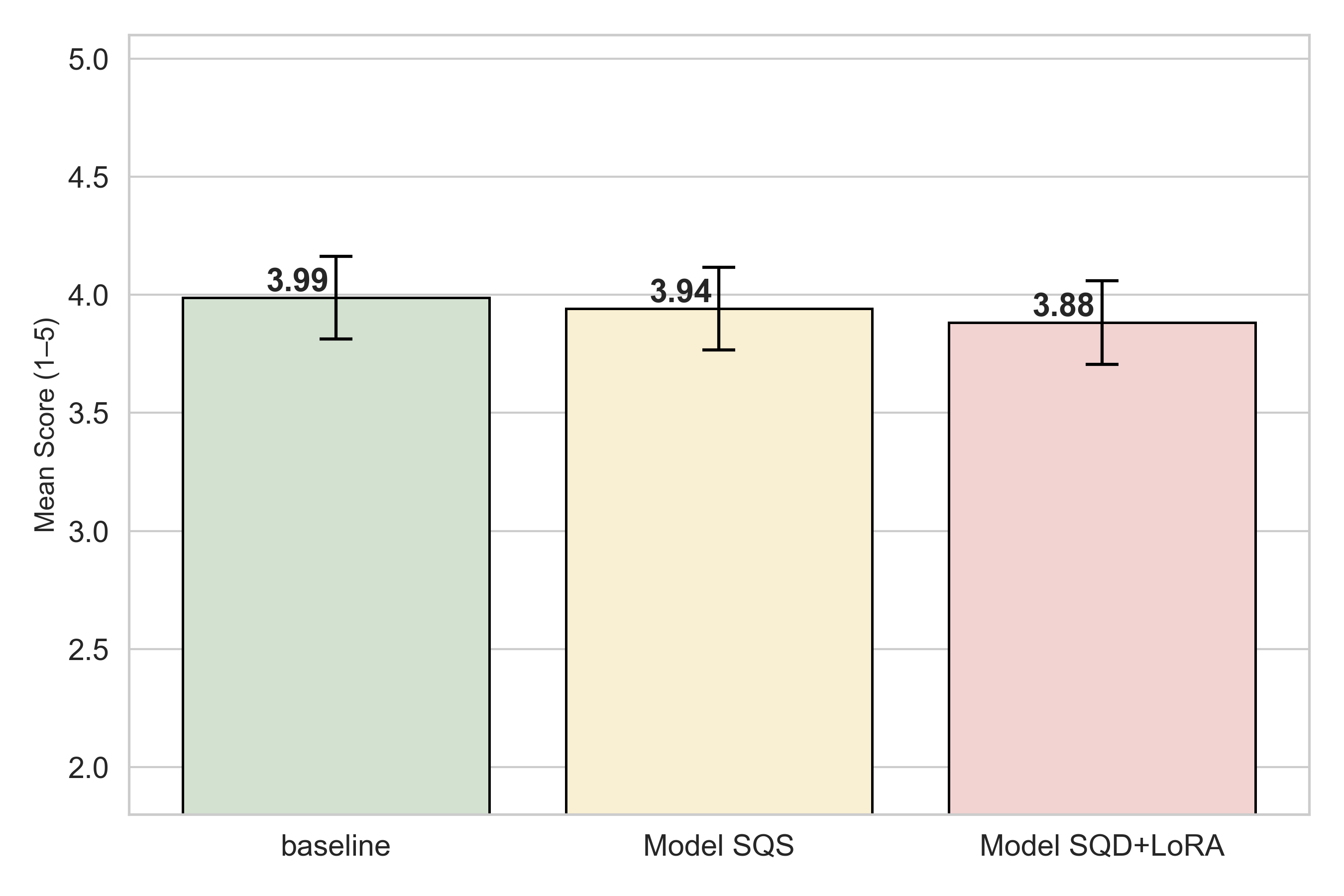}
        \vspace{-3mm}
    \caption{Subjective evaluation of mean user ratings (1–5 scale) in W8A8 for the full-precision baseline (baseline), the fastest variant (Model SQS), and the best-performing model (Model SQD+LoRA).}
    \label{fig:subjective_scores}
    \vspace{-3mm}
\end{figure}

Based on 100 ratings per model (300 in total), the full-precision baseline attained the highest mean score of 3.99. The W8A8-quantized SQS variant followed closely at 3.94, while the SQD + LoRA configuration achieved 3.88. This suggests that both quantized variants preserve perceptual quality remarkably well. In particular, the quantized SQS variant maintains perceptual fidelity nearly indistinguishable from the full-precision baseline. Meanwhile, the SQD + LoRA model achieves the highest CLAP score and remains competitive on other, but exhibits slightly lower subjective ratings—suggesting that LoRA fine-tuning may enhance objective alignment more than perceptual quality.

\section{Conclusions and Future work}

In this paper, we conducted a comprehensive study of PTQ strategies for audio DiTs, with a focus on the trade-offs between static and dynamic calibration. We introduced two practical extensions: denoising-timestep-aware smoothing and LoRA to compensate for residual weight errors. Our results show that SQD, with or without LoRA, preserves generation quality across both 8-bit and 4-bit settings, closely matching the full-precision baseline on objective metrics. While static quantization works well at 8 bits, it deteriorates at 4 bits, underscoring the need for dynamic calibration. LoRA improves 8-bit performance but has a limited impact at lower precision. A key limitation of SQD is slower inference, driven by the cost of dynamic scaling. Subjective evaluations confirm that quantized models remain perceptually close to the baseline. Future work will explore faster implementations of dynamic quantization and full low-precision execution to further accelerate inference of DiT audio models.

\section{Acknowledgments}

We would like to thank Google for its support in making generative audio models more lightweight through their Award for Inclusion Research (AIR) Program.

\bibliographystyle{IEEEtran}
\bibliography{refs25}







\end{document}